# A smartphone-based vision simulator

*Pragathi Praveena[1*], Jobin J Kavalam[1†], Namita Jacob[2]*

[1] Department of Electrical Engineering, Indian Institute of Technology Madras, Chennai 600036, India
[*]ee11b123@ee.iitm.ac.in, [†]ee09b050@ee.iitm.ac.in
[2] Chetana Charitable Trust, Chennai 600036, India
namitaj@chetana.org.in

## Abstract

Simulators, as tools that can clearly bring out the effect of impairment, are invaluable in the design and development process of an assistive device. Simulators are vital in meeting high standards of accessibility. Described is our work on a smartphone-based vision simulator for diabetic retinopathy that is economic, portable, flexible and easy-to-use.

## 1 Introduction

Accessibility is increasingly considered important in the design of any product. The current design flow for making the design accessible includes incorporating a list of good design practices or guidelines followed by verifying compliance with regulations (often using an automatic validator).

The validators are rule checking systems. For example, evaluating the accessibility of a webpage involves checking if every graphic in a web page has an alternate text version. While these rules are targeted at the barriers that an individual with impairment may encounter, it is difficult for the designer to evaluate first-hand the effectiveness of any single accessibility feature. Nonetheless, being able to do that evaluation is important. Most of all, it gives the designer a level of confidence about the accessibility of the design and helps make an estimate of how much more to do. Secondly, there may be factors specific to the design that the guidelines do not cover.

Therefore, there is a strong case for simulators or perhaps a whole suite of them that can also be easily incorporated into the design flow. Our work on a vision simulator for diabetic retinopathy is based on this philosophy.

## 2 Vision simulator for diabetic retinopathy

As part of the course, "Sensory, Motor and Language Disorders", the first in a series on Assistive Technology offered as a minor at Indian Institute of Technology Madras, we designed a vision simulator for diabetic retinopathy.

Diabetic retinopathy (DR) is a microvascular complication of diabetes mellitus which currently affects more that 170 million persons worldwide and will affect an estimated 336 million by 2030. DR develops in more than 75% of the patients who have had diabetes mellitus for more than 20 years [1]. The current simulator is an Android Application that captures video from the phone's camera and displays it on the screen just as a person with diabetic retinopathy is likely to see.

### 2.1 Implementation details

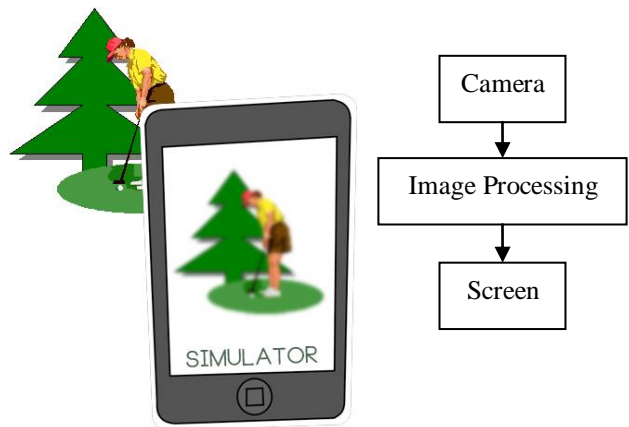

**Figure 1: Vision Simulator**

The simulator was built as an Android application using OpenCV (an open source computer vision library) to do the image processing. The application is compatible currently with Android 3.0 and above.

### 2.2 Simulations of stages of diabetic retinopathy

For normal vision, only a small oasis of clarity exists at the point of fixation and the scene grows progressively blurred at the periphery. On the other hand, when the same scene is viewed through a camera, there is no difference between the clarity of the centre and periphery. Hence, to reasonably simulate normal vision on the screen we introduced eccentric blurring. Values from prior studies [2] have been used to increasingly blur the image as we go away from the centre using a built-in blurring function of OpenCV.

DR is often associated with other complications resulting from the effect of diabetes on all retinal cell types [3]. We have included a few of the associated complications as well in the simulator. The progressive vision loss has been divided into four stages which approximately follow the ophthalmological classification.

The initial stage of DR is termed as non-proliferative (NPDR) and causes no vision loss. However, macular edema, an associated complication, is known to be a major cause of visual deterioration in patients with DR [4]. Macular edema causes loss of central vision simulated as a hazy spot in the centre of the scene. This has implications on the quality of communication and mobility (Figure 2). Using the simulator, some of the problems we identified included difficulty in making out facial expressions of an individual, identifying objects or people in crowded situations, detecting curbs and correctly recognising road signs and traffic signals.

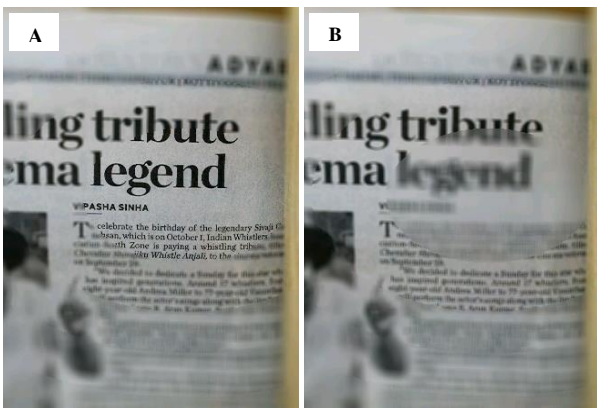

**Figure 2: (A) Original scene (B) Screenshot of the simulator showing degeneration of central vision affecting reading. Notice the blurred "legend".**

Studies have shown that patients with DR have an increased incidence of acquiring tritanopia (blue-yellow colour blindness) [5]. Moreover, the treatment of DR with laser photocoagulation is shown to produce permanent tritanopia [6]. Using simulation matrices [7], in stage 2 we have simulated blue-yellow colour deficit with different severities. The major implication of this acquired deficit will be in home glucose monitoring which depends on correctly interpreting the blues and yellows in colour-reagent test strips and may involve comparing the strip colour to a colour concentration chart (Figure 3).

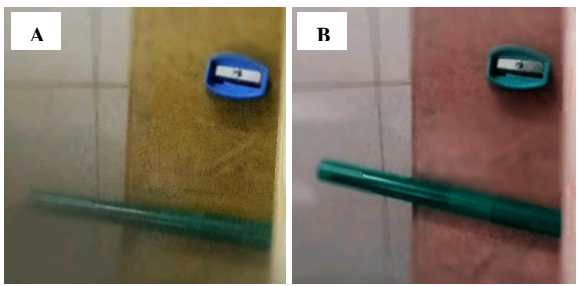

**Figure 3: (A) Original scene (B) Screenshot of the simulator showing blue-yellow colour deficiency. Notice that the sharpener and the pen appear to have the same colour.**

In NPDR, the blood vessels in the retina are weakened and tiny bulges called microaneurysms protrude from their walls. To improve blood circulation in the retina, new fragile blood vessels may form on the retinal surface which can leak blood into the vitreous. Patients may see a few specks of blood in their field of vision and some clouding which tends to disappear gradually. These effects are simulated in stage 3. The next stage of DR – proliferative diabetic retinopathy (PDR) is associated with severe vision loss and black patches that appear in the field of vision (Figure 4).

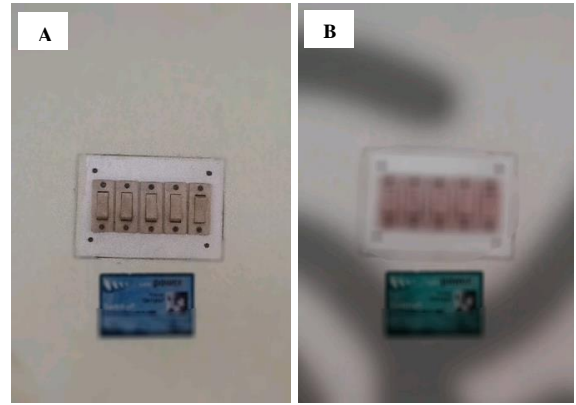

**Figure 4: (A) Original scene (B) Screenshot of the simulator showing a row of switches. With reduced contrast sensitivity and acuity along with the black patches, it is difficult to make out which switch is on from a distance.**

## 3 Discussion

Using a mobile application on smartphone has a big advantage in terms of portability. The simulator can be used on the spot to see how vision loss affects communication, learning and mobility. For a designer, this is important in ensuring the completeness of the design. As described earlier bringing in accessibility is often about fulfilling a checklist of good practices. However, these generic recommendations do not cover all situations and in many cases there may be issues specific to the problem at hand. An example of this will be the home glucose monitoring mentioned earlier. Such considerations are often easily overlooked and this is where the rule based approach becomes cumbersome. Instead, if the measuring process is rehearsed with the simulator in place, there are high chances that the difficulties will become apparent.

We can never be sure if a simulator correctly reproduces the actual vision loss. The process of development will invariably involve a lot of trial and error. The ease with which the code can be changed is an advantage over hardware simulators.

Development of an assistive device involves asking a patient to use a prototype repeatedly and give feedback. This is not only time-consuming, but may also turn out to be frustrating for both the designer and patient. To make the product more generic and cater to a wider section of people would also need such trials to be repeated on many patients. When products are designed by able-bodied individuals who may not fully understand the limitations or strengths of the user's disability, a simulator provides an inexpensive and versatile method of recognising it in the preliminary stages. Parameters can be modified to reasonably model different scenarios. Further, simulators can model future scenarios which would be especially useful in designing products for degenerative disorders.

Our vision simulator is quite effective in showing the effect of visual deterioration due to DR on communication, learning and mobility. However, each stage is only an average case in the progression of DR and may not exactly match an individual's experience. Also, there might be situations where discrete steps are not sufficient to evaluate a design's effectiveness. We have observed that the image processing for eccentric blurring causes a delay which makes the simulation sluggish. Being able to calibrate the simulation with an individual patient's medical record which has details of his/her acuity, contrast sensitivity, colour vision, retinal scans etc. will be a useful, albeit challenging addition to the simulator.

# 4 Conclusion

The key challenge in the development of products for people with disabilities as well as in making existing designs accessible is the lack of understanding of the disability itself. With the help of a portable simulator, designers can put themselves in the shoes of the user. Further, the simulator's flexibility makes it easy to evaluate the design's effectiveness over a range of disabilities and its varying degrees.

Thus, wide adoption of simulators will go a long way in setting high standards of accessibility. It will also make for a common benchmark against which new designs can be compared.

Simulators on smartphones can serve well not only to help the designer but also aid the person with disability or the caregiver to make suitable lifestyle changes to enable independent living.

## Acknowledgements

We thank Ranjani Srinivasan (Dept. of Engineering Physics) and Umesh Surepalli (Dept. of Electrical Engineering), for their contribution to the project.